\documentclass[letterpaper]{mc2021}

\usepackage{subcaption}
\usepackage{booktabs}
\usepackage{nth}
\usepackage[ruled,vlined,noend]{algorithm2e}
\usepackage{placeins}
\usepackage{physics}

\usepackage{tabls}
\usepackage{cites}
\usepackage{epsf}
\usepackage{appendix}
\usepackage{ragged2e}
\usepackage[top=1in, bottom=1in, left=1in, right=1in]{geometry}
\usepackage{enumitem}
\setlist[itemize]{leftmargin=*}
\newcommand{\bOmega}{\boldsymbol{\Omega}}

\newcommand{\refeqn}[1]{Eq.~\eqref{eqn:#1}}
\newcommand{\refeqns}[1]{Eqs.~\eqref{eqns:#1}}

\title{A NEW SCHEME FOR SOLVING HIGH-ORDER \\ DG DISCRETIZATIONS OF THERMAL RADIATIVE TRANSFER \\ USING THE VARIABLE EDDINGTON FACTOR METHOD}
\author{%
  \textbf{Ben C. Yee$^1$, Samuel S. Olivier$^{1,2}$,} \\ \textbf{Ben S. Southworth$^3$, Milan Holec$^1$, and Terry S. Haut$^1$} \\
  $^1$Lawrence Livermore National Laboratory \\
  7000 East Avenue, Livermore, CA 94550 \\ 
\\
  $^2$Applied Science and Technology, University of California, Berkeley, CA 94708 \\
     \\
  $^3$Los Alamos National Laboratory, Los Alamos, NM, 87545 \\
     \\
  \url{yee26@llnl.gov}, \url{solivier@berkeley.edu}, \\ \url{ben.s.southworth@gmail.com}, \url{holec1@llnl.gov}, \url{haut3@llnl.gov}
}
\newcommand{\authorHead}{Yee, Olivier, et al.}
\newcommand{\shortTitle}{High-order TRT with VEF}
\begin{document}
\maketitle
\justify 

\begin{abstract}
  We present a new approach for solving high-order thermal radiative transfer (TRT) using the Variable Eddington Factor (VEF) method (also known as quasidiffusion).  Our approach leverages the VEF equations, which consist of the first and second moments of the $S_N$ transport equation, to more efficiently compute the TRT solution for each time step.  The scheme consists of two loops -- an outer loop to converge the Eddington tensor and an inner loop to converge the iteration between the temperature equation and the VEF system.  By converging the outer iteration, one obtains the fully implicit TRT solution for the given time step with a relatively low number of transport sweeps.  However, one could choose to perform exactly one outer iteration (and therefore exactly one sweep) per time step, resulting in a semi-implicit scheme that is both highly efficient and robust.  Our results indicate that the error between the one-sweep and fully implicit variants of our scheme may be small enough for consideration in many problems of interest.
\end{abstract}

\section{INTRODUCTION} 
We are interested in solving $S_N$ thermal radiative transfer (TRT) on high-order, curved meshes with arbitrary-order discontinuous Galerkin (DG) spatial discretizations. 
Our motivation stems from the increasing popularity of high-order spatial discretizations and meshes in hydrodynamics simulations.
For Lagrangian and arbitrary Lagrangian-Eulerian hydrodynamics, high-order methods have been shown to provide greater robustness (especially if mesh distortions are present), improved symmetry preservation, and stronger scaling compared to low-order, straight-mesh methods~\cite{langer2015performance,dobrev2012high,anderson2018high}.
In many applications of interest (inertial confinement fusion, astrophysical phenomena), hydrodynamics and TRT are tightly coupled, and solving both physics on the same spatial mesh is desirable for accuracy, efficiency, and robustness.
The standard alternative -- mapping hydrodynamic quantities onto a low-order refined mesh  for TRT -- hinders the simulation stability and increases the number of unknowns, compounding the aforementioned difficulties of numerically modeling TRT.

In this work, we present an efficient approach for solving the time-dependent TRT equations in the context of DG on curved meshes.
Our new algorithm is heavily inspired by the one developed by Paul Nowak and others in the Teton library for solving TRT on unstructured straight-edged meshes using an upstream corner balance spatial discretization.
\cite{brunner2020nonlinear} provides insight into the details of Teton's nonlinear iteration scheme.
The work described in this abstract leverages previous work in the Variable Eddington Factor (VEF) method (also known as quasidiffusion)~\cite{olivier2017variable,gol1964quasi} and flux-fixups for DG discretizations of transport~\cite{yee2020quadratic}.
VEF is a well-known approach for accelerating source iteration and can produce solutions efficiently in both optically thin and optically thick regimes.
However, high-order DG discretizations of TRT can produce negative angular intensities, which are problematic for the evaluation of the Eddington tensor.
In~\cite{yee2020quadratic}, it is shown that various sweep-compatible and balance-preserving fixups can be applied to the angular intensity to remedy this issue without degrading the convergence properties of VEF.


In the following sections, we provide some light background for the method, provide an overview of our new algorithm, and demonstrate the scheme on two test problems -- a modified version of the MMS problem from \cite{brunner2006development} and a simplified 2D-XY version of a test problem from the National Ignition Facility (NIF).
We focus on the important aspects of our algorithm in this conference paper and defer the remaining details to a future journal article.
The preliminary results indicate that our new algorithm has the potential to solve difficult TRT problems robustly and efficiently.
Here, ``difficult'' means heterogeneous in optical thickness (having both very thin and thick regions) and plagued by ray effects.
One particularly intriguing aspect of this algorithm is that it suggests the possibility of only needing one transport sweep per TRT time step. 

We note that there are two related M\&C submissions from the coauthors of this work.
Olivier describes a new DG-based discretization for VEF in \cite{olivier2021mc}.
This new discretization is of particular interest because the results of \cite{olivier2021mc} demonstrate that the discretized equations can be solved efficiently, regardless of order, mesh size, and penalty parameter.
(The work in this paper uses the mixed finite element formulation described in~\cite{olivier2019high} for VEF, but we will switch to the new DG-based formulation in future work.)
In \cite{holec2021mc}, Holec presents an alternative approach for high-order TRT based on diffusion synthetic acceleration (DSA) rather than VEF.
Like the VEF-based approach described in this summary, the nonlinear DSA (NDSA) method also offers the possibility of only performing one sweep per time step.

\section{THEORY}

The grey, discrete ordinates (S$_N$) TRT equations are given by:
\begin{subequations}
\begin{gather}
   \frac{1}{c} \pdv{I_d}{t} + \boldsymbol{\Omega}_d \cdot \nabla I_d + \sigma I_d = \sigma B(T) \,, \quad \quad d = 1, \ldots, D \label{eqn:TRT_trans}\,, \\
   \rho c_v \pdv{T}{t} + 4\pi \sigma B(T) = \sigma \varphi \label{eqn:TRT_temp} \,.%
\end{gather} \label{eqns:TRT}%
\end{subequations}%
Here, the notation is standard:
$I$ is the angular intensity, $d$ is the index of discrete ordinate direction, $\varphi$ the scalar intensity, $T$ is the material electron temperature, and $B(T) \equiv ac T^4 / (4\pi)$.

Typically, one obtains $\varphi$ in \refeqn{TRT_temp} by taking a weighted sum $I_d$.  With VEF, however, $\varphi$ is obtained by solving a system consisting of the first two angular moments of \refeqn{TRT_trans}:\\
\begin{minipage}{0.5\textwidth}
\begin{subequations}
\begin{gather}
	\frac{1}{c} \pdv{\varphi}{t} + \nabla \cdot \boldsymbol{J} + \sigma \varphi = 4\pi \sigma B(T) \,, \label{eqn:VEF_phi} \\
	\frac{1}{c} \pdv{\boldsymbol{J}}{t} + \nabla \cdot \left( E \varphi \right) + \sigma \boldsymbol{J} = 0 \,, \label{eqn:VEF_J}%
\end{gather} \label{eqns:VEF}%
\end{subequations}%
\end{minipage}%
\begin{minipage}{0.5\textwidth}
\begin{equation}
	E \equiv \cfrac{\sum\limits_d w_d \bOmega_d \otimes \bOmega_d I_d}{\sum\limits_d w_d I_d} \,.
\end{equation}%
\end{minipage}\\
Here, $\boldsymbol{J}$ is the radiation current (or flux), and $E$ is the Eddington tensor.
One unique aspect of VEF is that \refeqn{TRT_trans} and \refeqns{VEF} are discretized independently of each other and do not have to be consistent (i.e., $\varphi \ne \sum\limits_d w_d I_d$).
A backward Euler discretization is applied for the time variable in this work.
However, it is straightforward to generalize our algorithm to other implicit or semi-implicit time discretization techniques.
Further details regarding VEF, especially in the context of finite element discretizations, can be found in~\cite{olivier2017variable}, \cite{olivier2021mc}, and \cite{olivier2019high}.
\FloatBarrier

\section{METHOD} 
\begin{algorithm}[tb]
\SetAlgoLined
 Initialize $\varphi$, $T$, $E$ from previous time step\;
 $T$ = SolveNonlinearTemperatureEquation($\varphi$)\;
 \While{Outer tolerance not satisfied}{
 	(if not first outer) $E$ = TransportSweepToUpdateEddingtonTensor($\varphi$,$T$)\;
    \While{Inner tolerance not satisfied}{
       $\varphi$ = SolveLinearizedVEFEquations($E$,$T$)\;
 	   $T$ = SolveNonlinearTemperatureEquation($\varphi$)\;
    }
 }%
 \caption{TRT-VEF algorithm for one time step} \label{alg:TRTVEF}%
\end{algorithm}%

\begin{figure}[htb]
  \centering
  \includegraphics[width=0.75\linewidth]{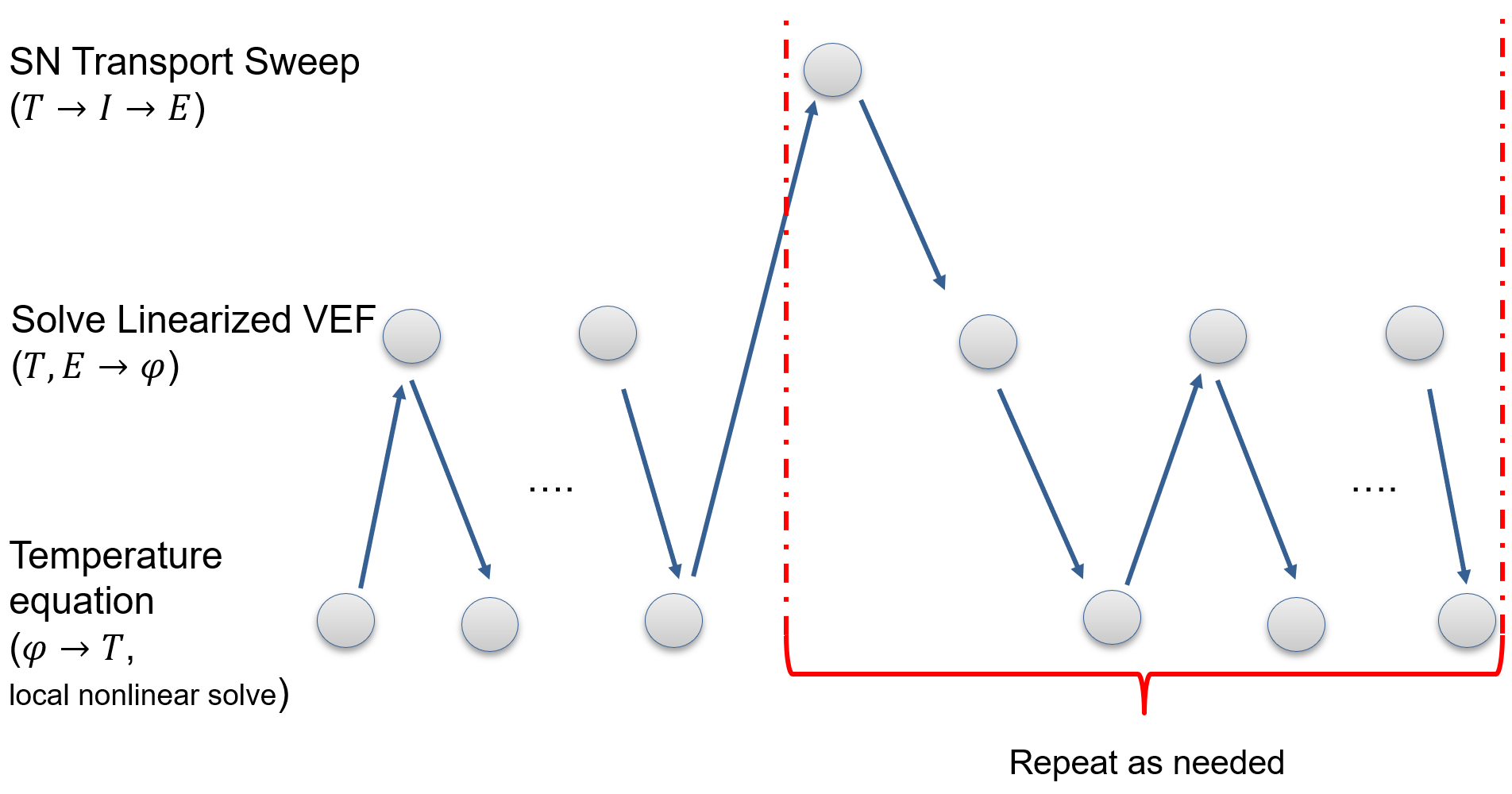}%
  \caption{Visualization of the TRT Algorithm in Algorithm~\ref{alg:TRTVEF}.} \label{fig:algo}%
\end{figure}

Algorithm~\ref{alg:TRTVEF} provides a succinct description of our proposed TRT-VEF algorithm.
The basic idea is that we iterate between three sets of equations: the transport equation (\refeqn{TRT_trans}),  the VEF equations (\refeqns{VEF}), and the material temperature equation (\refeqn{TRT_temp}).
The latter two are much more tightly coupled, and it is beneficial to iterate between these two sets of equations frequently.
The transport equation only impacts the other two components via the Eddington tensor.
Because the Eddington tensor converges relatively quickly~\cite{olivier2017variable} and transport sweeps are computationally expensive, $E$ is not updated in the inner iteration.
Figure~\ref{fig:algo} provides a visualization of this iteration.

The linearized VEF system solved in the first part of each inner iteration is derived as follows.
First, we replace each instance of $B(T)$ in \refeqn{TRT_temp} and \refeqn{VEF_phi} using
\begin{equation}
	B(T) \to B(T^*) + \left.\frac{dB}{dT}\right|_{T^*} \left[ T - T^* \right] \,,
\end{equation}
where $T^*$ is the most recent temperature iterate.
Then, we substitute this into the time-discretized version of \refeqn{TRT_temp}, solve for $T$, and use the result to eliminate $T$ from the time-discretized, linearized version of \refeqn{VEF_phi}.
(\refeqn{VEF_J} remains unchanged.) 
More details will be provided in the full paper. 

Though the outer iteration can be repeated to converge to the fully implicit backward Euler solution, our preliminary results indiciate that it is possible to ``get away'' with only sweeping once per time step (two outer iterations in Algorithm~\ref{alg:TRTVEF}).
Doing so effectively produces a semi-implicit solution and can provide a reasonable compromise between robustness, accuracy, and computational burden.
We note that the transport sweep here is fully upwind (no lagged fluxes at domain boundaries), but we may study the potential benefits of loosening this restriction in future work.

Fixups are applied on all three levels.
During the transport sweep, the QPMP fixup from~\cite{yee2020quadratic} is used.
In the inner iterations, the QPZ fixup from~\cite{yee2020quadratic} is used to ensure that $\varphi$ and $T$ are above a small positive floor.
The 2$\times$2 VEF system (\refeqns{VEF}) is solved using GMRES with a block triangular preconditioner.
The inverse of the block triangular is applied using HypreBoomerAMG on an approximate Schur complement explicitly formed through the use of finite element mass lumping on the diagonal block of \refeqn{VEF_J}.
The convergence criterion for both the inner and outer loops is the space-integrated $L_2$ norm of the change in $T$ from iteration to iteration.
In future work, we may consider the convergence of other terms in the criteria (e.g., the residual, $E$, and/or $\varphi$).

We note that Algorithm~\ref{alg:TRTVEF} can be described nicely as an inexact Newton iteration with nonlinear elimination using the framework set out in~\cite{brunner2006development}.  This connection will be hashed out in detail when this conference paper is extended into a journal article.

\section{RESULTS AND DISCUSSION}%

\subsection{MMS Problem}
We first demonstrate our new scheme on a modified version of the 2D-RZ radiation diffusion MMS problem defined in~\cite{brunner2006development}.  We define the following 2D-XY versions of the solutions to $T$ and $\varphi$:
\begin{gather}
T(x,y,t) = T_0 \left[ 1 - \frac{e^{-\tau t}}{2} \cos \left( \omega x \right) \cos \left( \omega y \right) \right] \label{eqn:T_mms} \\
\bar{T}_{\text{rad}}(t) = T_0 \left[ 1 + \frac{e^{-\tau t}}{2} \right] \\
\varphi(x,y,t) = a c \bar{T}_{\text{rad}}^4\left[ 1 + \frac{e^{-\tau t}}{2} \cos \left( \omega x \right) \cos \left( \omega y \right) \right] \label{eqn:phi_mms} \,.%
\end{gather}%
To make this MMS problem a transport problem, we define $I$ as follows:
\begin{gather}
\theta(\bOmega) = \left[ \frac{1}{2} + \frac{3}{2}\Omega_z^2 \right] \,,  \quad \quad \theta_1(\bOmega) = \frac{1}{3} \Omega_x \Omega_y \,, \\
\varphi_1(\bOmega) =  a c \bar{T}_{\text{rad}}^4  \frac{e^{-\tau t}}{2} \sin \left( \omega x \right) \sin \left( \omega y \right)  \,, \\
I(\bOmega, x, y, t) = \frac{1}{4\pi} \left[ \varphi(x,y,t) \theta\!\left(\bOmega\right)  + \varphi_1(x,y,t) \theta_1\!\left(\bOmega\right)  \right] \label{eqn:psi_mms} %
\end{gather}%
%
We note that integrating $\theta(\bOmega)$ and $\theta_1(\bOmega)$ over all directions yields $4\pi$ and 0, respectively.

\begin{figure}[htb]
    \begin{subfigure}[t]{0.3\textwidth}
        \includegraphics[width=1.2\textwidth]{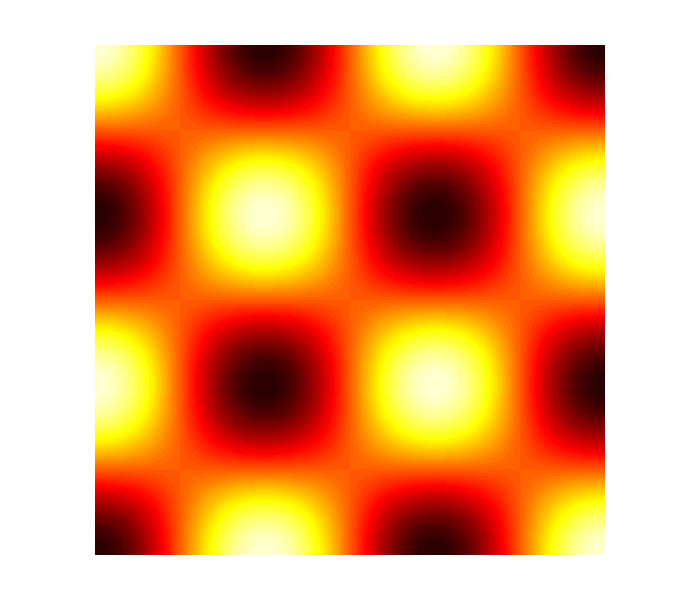}
        \caption{Exact MMS solution}
    \end{subfigure}
    \begin{subfigure}[t]{0.3\textwidth}
        \includegraphics[width=1.2\textwidth]{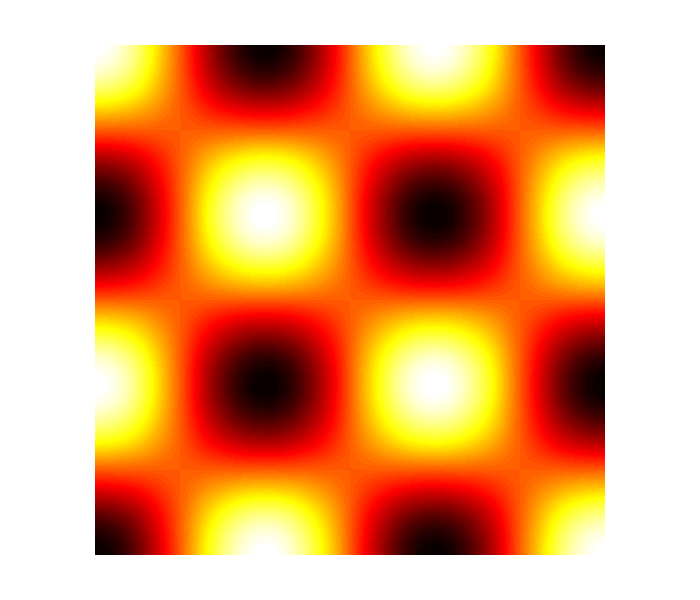}
        \caption{Fully implicit}
    \end{subfigure}
    \begin{subfigure}[t]{0.3\textwidth}
        \includegraphics[width=1.2\textwidth]{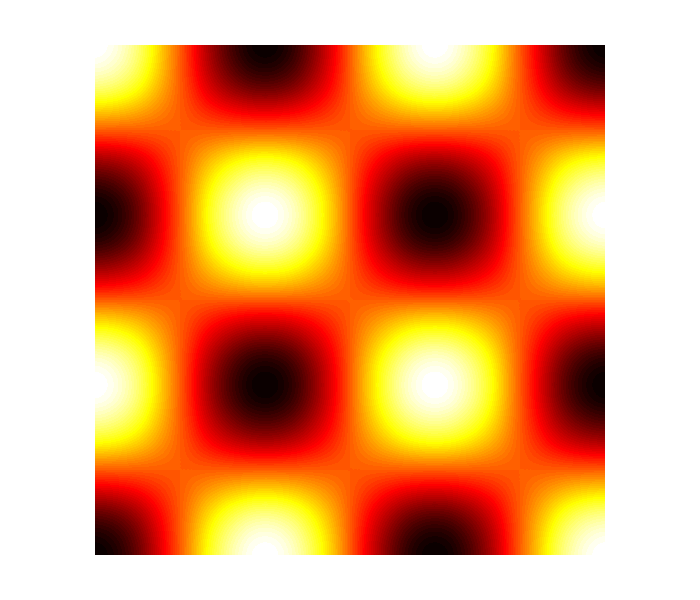}
        \caption{One sweep per $\Delta t$}
    \end{subfigure}%
    \begin{subfigure}[b]{0.09\textwidth}
    	\centering
    	\includegraphics[width=\textwidth]{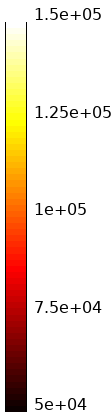}
    \end{subfigure}%
    \caption{$T(x,y)$ at $t_{final}$ for the MMS problem with $p = 2$ and 1 refinement.}%
    \label{fig:results_mms}%
\end{figure}

The values for the constants are defined in Table I of \cite{brunner2006development}.  
Following \cite{brunner2006development}, our domain is a square spanning $0 \le x,y \le 3\pi\omega^{-1}$, and the simulation is run from $t_{\text{start}} = 0$ until $t_{\text{final}} = \left(10 \tau \right)^{-1}$.  The coarsest discretization used a uniform square spatial grid with elements of size $\Delta x_{\text{coarse}} = 3\pi\omega^{-1}/5$ and a time step size of $\Delta t = t_{\text{final}}/2$. 
The spatial discretization and time step are refined together: with each refinement, we halve the spatial grid spacing (quadrupling the number of spatial elements) and we divide the time step by a factor of $2^{p+1}$ where $p$ is the degree of the polynomials used in the DG representation of $\varphi$ and $T$.
(DG is a $(p+1)$-th-order scheme while implicit Euler is a \nth{1}-order scheme.)
Because of this, we expect that the error should decrease by a factor of $2^{p+1}$ with each refinement.
An $S_{18}$ level-symmetric quadrature set is sufficient to neglect the angular error and observe the aforementioned convergence rate.
For this problem, the inner and outer tolerances are set to $10^{-12}$ and $10^{-14}$, respectively, so that the iteration error does not affect the convergence study.  (The outer tolerance is not used for the one-sweep runs.)
Snapshots of the solution are shown in Figure~\ref{fig:results_mms}.
\begin{figure}[tb]
	   \centering
    \begin{subfigure}[t]{0.49\textwidth}
	   \centering
        \includegraphics[width=\textwidth]{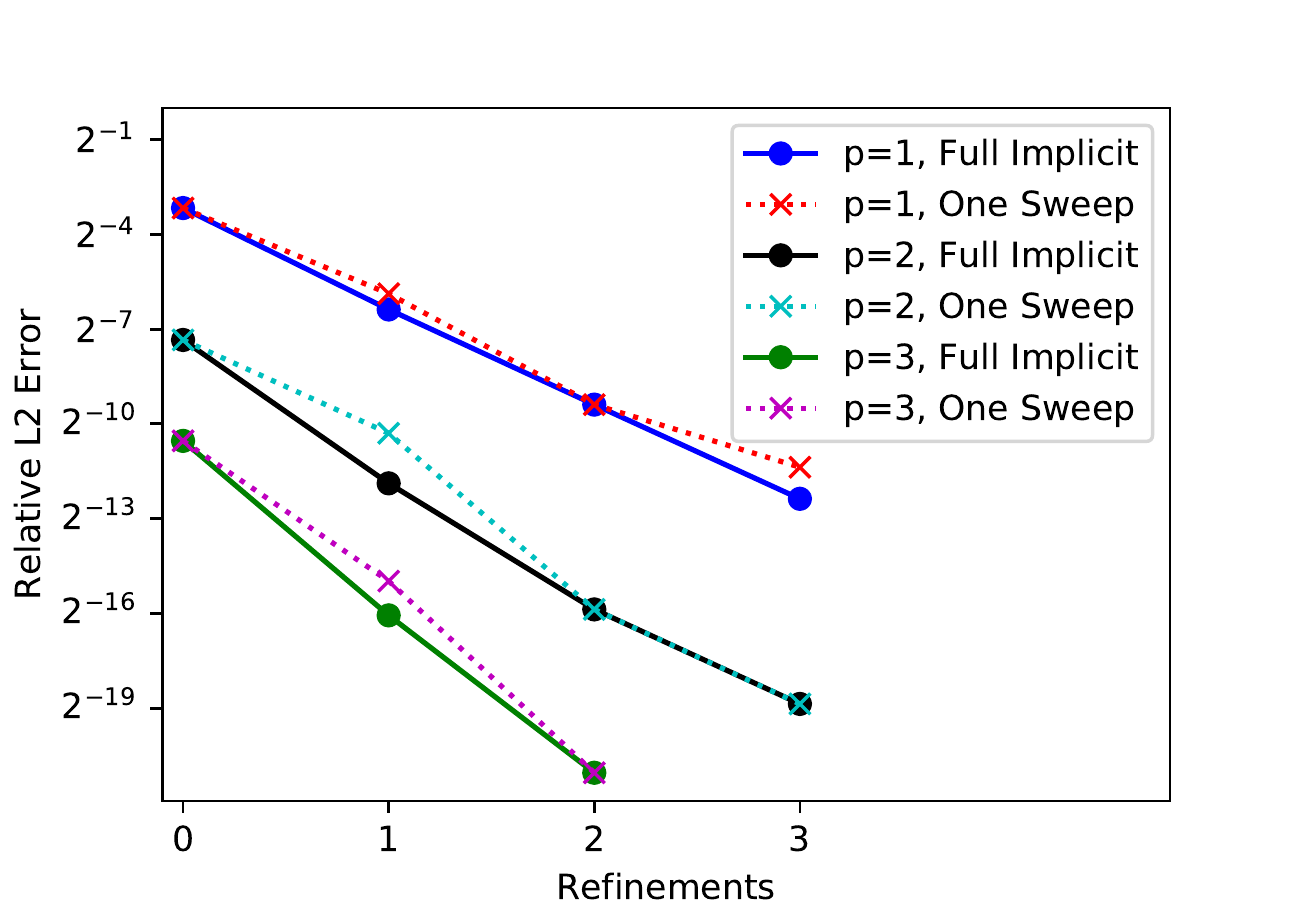}
        \caption{Error for $T$}
    \end{subfigure}
    \begin{subfigure}[t]{0.49\textwidth}
	   \centering
        \includegraphics[width=\textwidth]{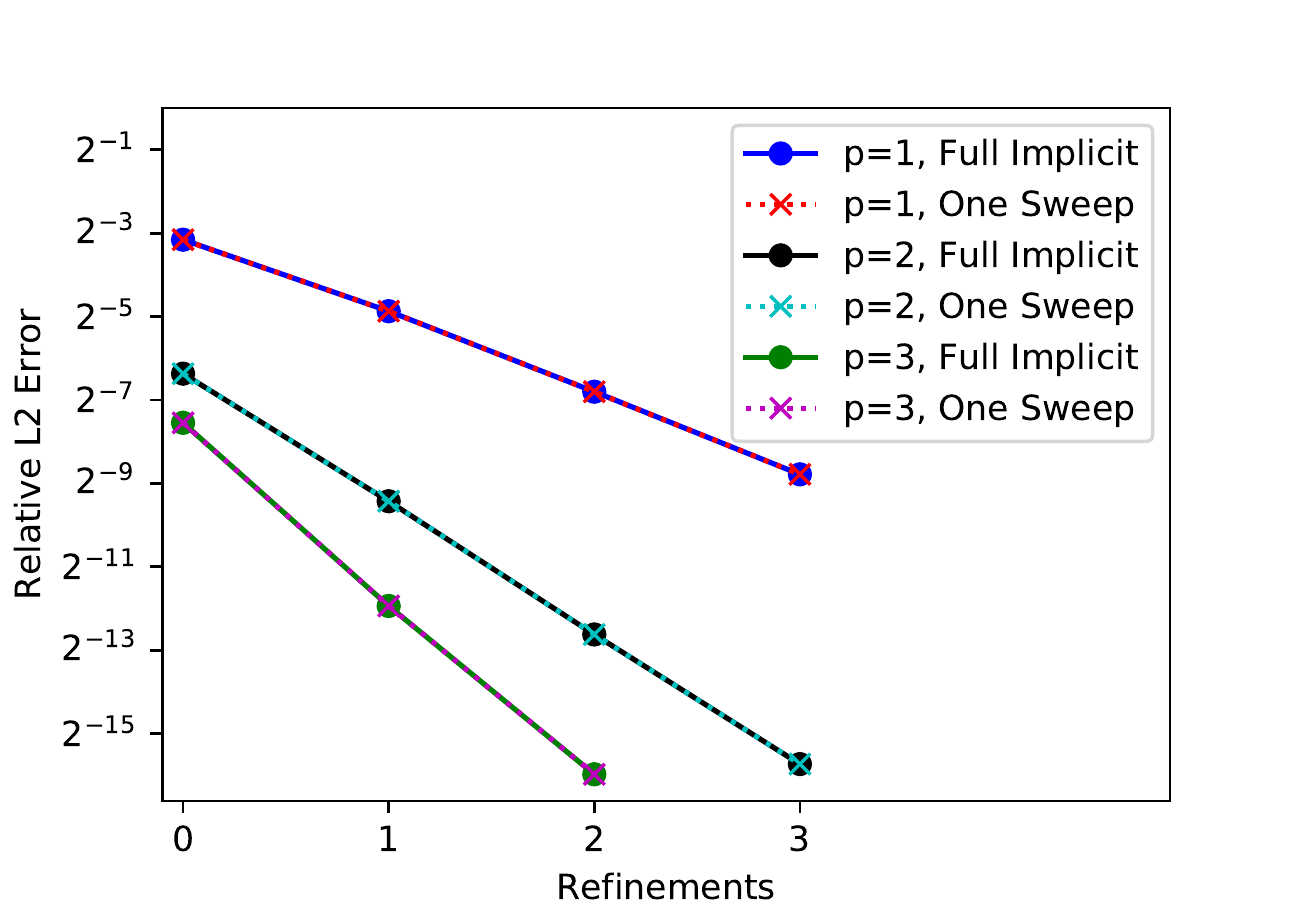}
        \caption{Error for $\varphi$}
    \end{subfigure}
        \caption{Relative space- and time-integrated $L_2$ errors for the MMS problem.} \label{fig:error_mms}%
\end{figure}%
\begin{table}[tb]
  \centering
  \caption{\bf Estimated orders of convergence for each line in Figure~\ref{fig:error_mms}.}
  \label{table:order_mms} 
  \begin{tabular}{c|cc|cc|cc}
  \toprule
  Order & \multicolumn{2}{c|}{1} & \multicolumn{2}{c|}{2} & \multicolumn{2}{c}{3} \\
  Variable & $T$ & $\varphi$ & $T$ & $\varphi$ & $T$ & $\varphi$ \\ \midrule
  Fully Implicit 		   & 3.1 & 1.9 & 4.0 & 3.1 & 5.3 & 4.2 \\
  One Sweep per $\Delta t$ & 2.8 & 1.9 & 4.0 & 3.1 & 5.3 & 4.2 \\
  \bottomrule%
  \end{tabular}%
\end{table}%
\begin{table}[tb]
  \centering
  \caption{\bf Average iterations per $\Delta t$ for the $p = 2$ case of the MMS problem.}
  \label{table:its_mms} 
  \begin{tabular}{c|cc|cc|cc|cc}
  \toprule
  Refinements & \multicolumn{2}{c|}{0} & \multicolumn{2}{c|}{1} & \multicolumn{2}{c|}{2}  & \multicolumn{2}{c}{3} \\
   & Outers$^\dagger$ & Inners & Out. & In. & Out. & In. & Out. & In.\\ \midrule
  Fully Implicit & 3 & 9 & 2 & 6 & 1 & 4 & 1 & 4\\
  One Sweep per $\Delta t$ & 1 & 5 & 1 & 4 & 1 & 4 & 1 & 4\\
  \bottomrule
  \end{tabular} \\ %
  ${}^\dagger$Outer iterations with a transport sweep (excludes the first outer in Algorithm~\ref{alg:TRTVEF}).%
\end{table}%

Space- and time-integrated errors for different $p$ values are shown in Figure~\ref{fig:error_mms} for several refinement levels, and estimated orders of convergence from a least-squares line fit are provided in Table~\ref{table:order_mms}.
In these plots, we see fairly small differences between the errors of the fully implicit and one sweep approaches.
More importantly, both sets of curves converge at the approximately same rate, thus demonstrating that the one-sweep approach maintains the same asymptotic 
accuracy as the fully-implicit approach.
In Table~\ref{table:its_mms}, we see that, for problems with sufficient resolution in space and time, the ``fully implicit'' approach only requires one outer iteration per time step, making it equivalent to the one-sweep approach.
Interestingly, we are seeing that $T$ consistently converges at approximately order $p+2$ -- one order faster than $\varphi$ -- despite the fact that $T$ is only represented using degree-$p$ polynomials.
We are unsure at the moment why this is this case, and we will seek an explanation in future work.

\subsection{Half-Hohlraum Problem}

\begin{figure}[tb]
    \begin{subfigure}[t]{0.5\textwidth}
	   \centering
        \includegraphics[scale=0.3]{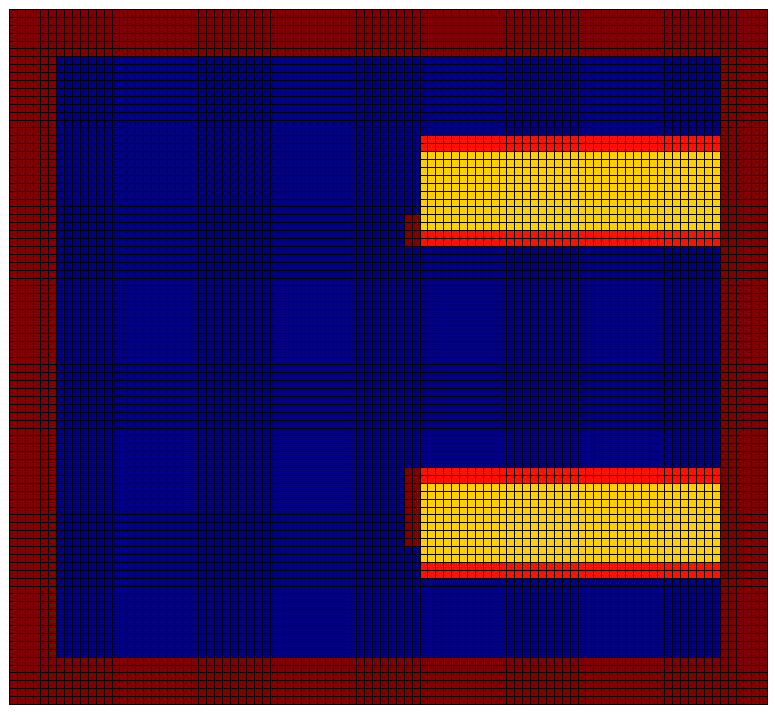}
        \caption{Materials: Au wall (dark red), Ta wall (light red),\\ Si foam (yellow), He fill (blue)}
    \end{subfigure}
    ~
    \begin{subfigure}[t]{0.5\textwidth}
	   \centering
        \includegraphics[scale=0.3]{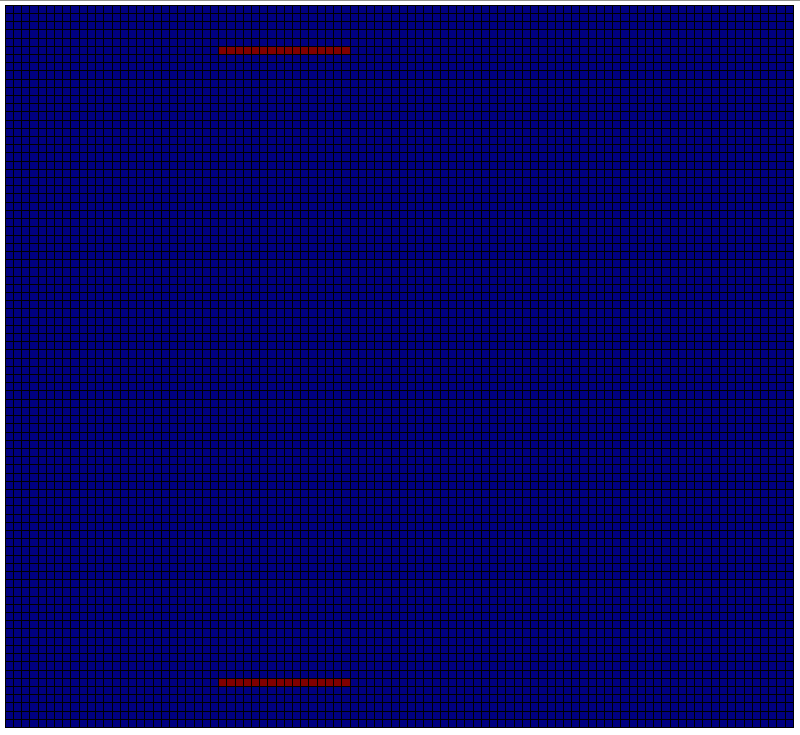}
        \caption{Initial conditions (red $=$ hot zones, blue $=$ cold zones)}
    \end{subfigure}%
    \caption{2-D (X-Y) grey model of the half-haulraum problem ~\cite{fhr}.} \label{fig:problem}%
\end{figure}

Next, we demonstrate some preliminary results on a 2D-XY model of a half-hohlraum inspired by radiation hydrodynamics studies from the NIF Sonoma campaign, with a focus on heat-wave propagation in the throttled foam channels~\cite{fhr}.
Figure~\ref{fig:problem} shows the spatial mesh used as well as the breakdown of the regions for the materials and the initial condition.
The spatial domain is 96 elements wide by 88 elements tall; each element is a square of length %
\begin{equation}
	\Delta = \frac{0.623}{88} \text{ cm} \,. \label{eqn:Delta} %
\end{equation}%
\begin{figure}[tb]
    \begin{subfigure}[t]{0.23\textwidth}
        \includegraphics[width=1.05\textwidth]{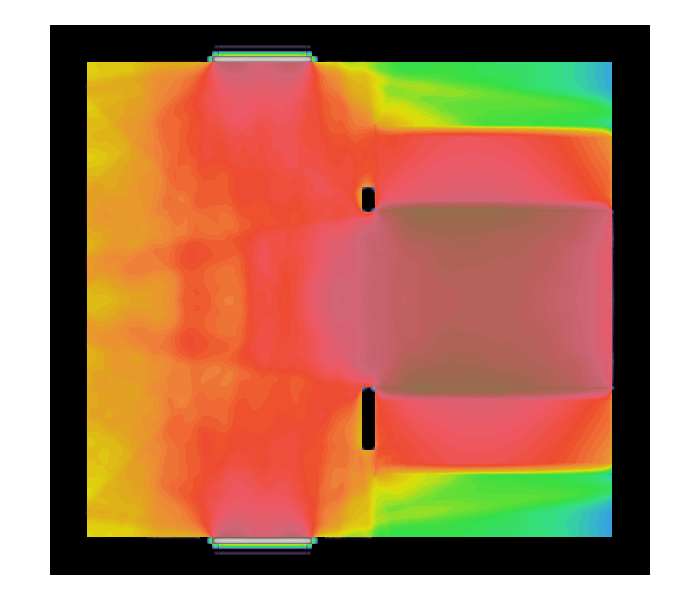}
        \caption{Reference solution \\ (10x more time steps)}
    \end{subfigure}
    ~
    \begin{subfigure}[t]{0.23\textwidth}
        \includegraphics[width=1.05\textwidth]{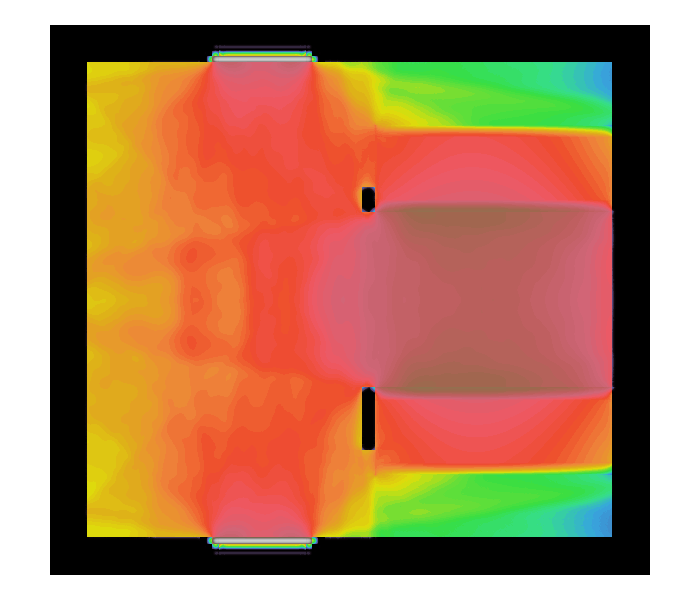}
        \caption{Fully implicit}
    \end{subfigure}%
    \begin{subfigure}[t]{0.23\textwidth}
        \includegraphics[width=1.05\textwidth]{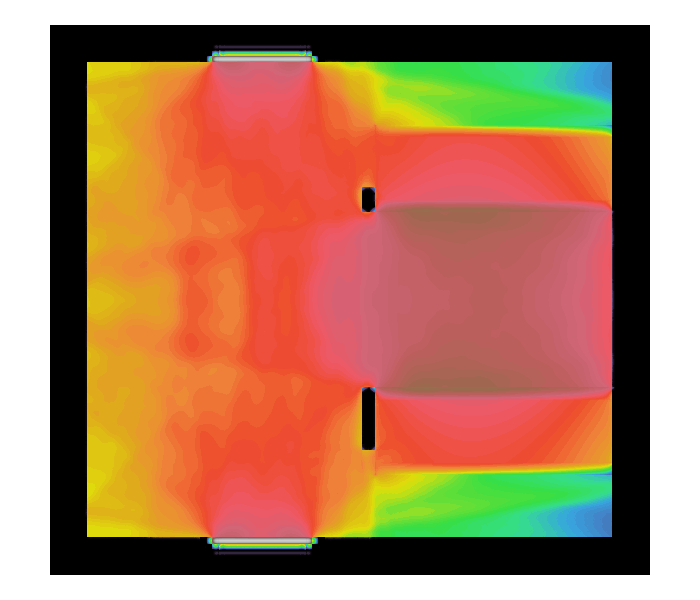}
        \caption{One sweep per $\Delta t$}
    \end{subfigure}%
    \begin{subfigure}[t]{0.23\textwidth}
        \includegraphics[width=1.05\textwidth]{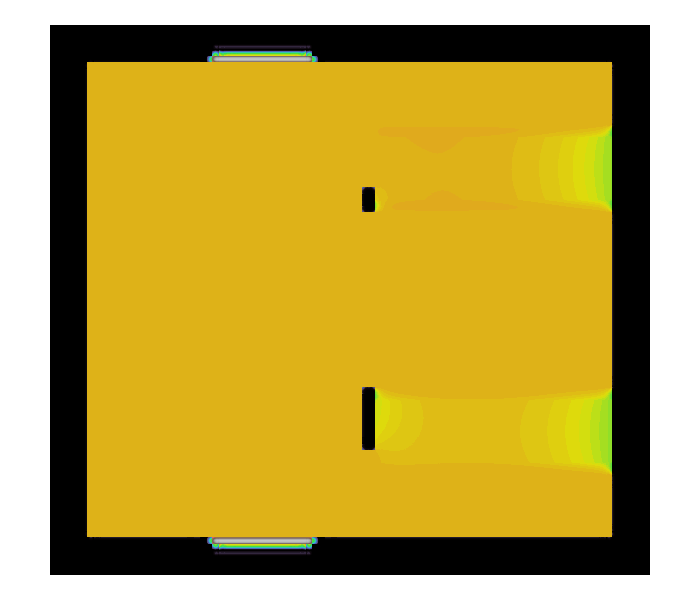}
        \caption{Diffusion ($E = \frac{1}{3}I$)}
    \end{subfigure}%
    \begin{subfigure}[b]{0.065\textwidth}
    	\centering
    	\includegraphics[width=\textwidth]{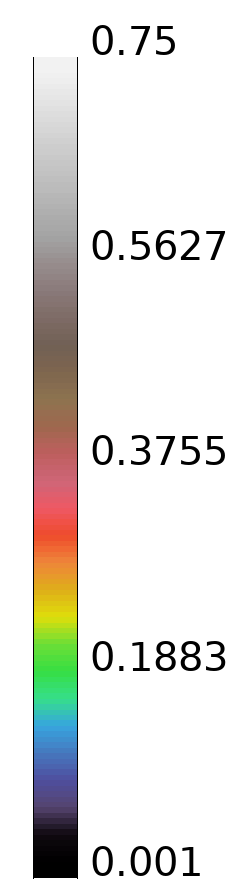}
    \end{subfigure}
    \caption{$T(x,y)$ at time step 50 ($t = 3$ ns) for the half-hohlraum problem.}%
    \label{fig:results}%
\end{figure}%
\begin{figure}[tb]%
\begin{minipage}{0.5\textwidth}
  \captionsetup{type=table} 
  \centering
  \caption{\bf Total iteration counts at $t = 3$ ns.}
  \label{table:results} 
  \begin{tabular}{ccccc}
  \toprule
   & Outers$^\dagger$ & Inners\\ \midrule
  Reference & 3020 & 9860 \\
  Fully Implicit&  860 & 5688 \\
  One Sweep per $\Delta t$ & 50 & 1273 \\
  Diffusion & 0 & 759 \\
  \bottomrule%
  \end{tabular}\\ \vspace{4pt} 
  ${}^\dagger$Outer iterations with a transport sweep (excludes the first outer in Algorithm~\ref{alg:TRTVEF}).%
\end{minipage}%
\begin{minipage}{0.5\textwidth}
\captionsetup{type=figure}
	   \centering
        \includegraphics[width=\textwidth]{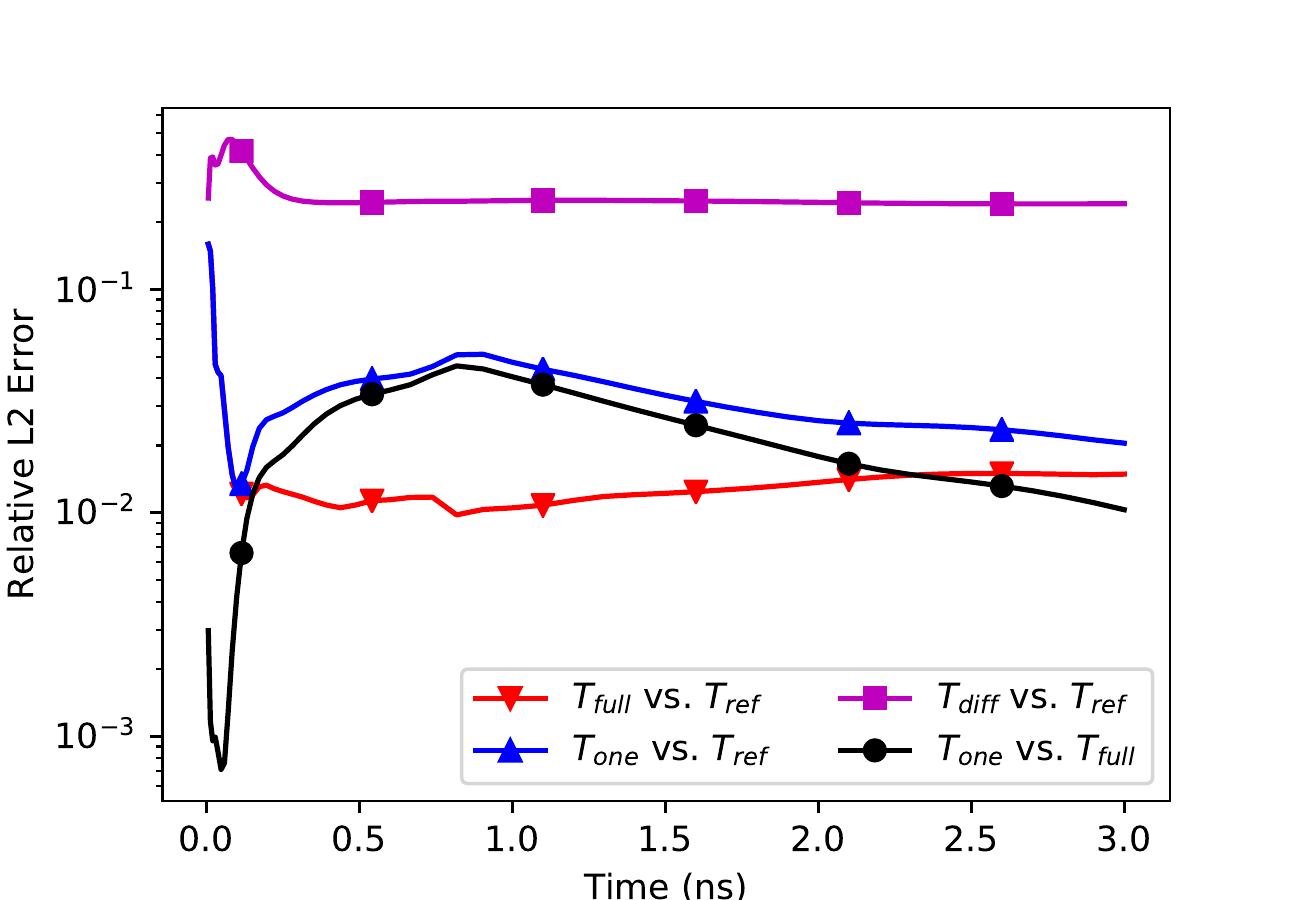}%
        \caption{Relative space-integrated $L_2$ errors \\ for $T$ up to time step 50 ($t = 3$ ns).} \label{fig:error}%
\end{minipage}%
\end{figure}%
A detailed description of the location of the material boundaries can be found in Appendix A.
(Though this problem has a Cartesian mesh, our code and algorithm also works for curved meshes.)
$\sigma$, $\rho c_v$, and initial conditions are given by:
\begin{subequations}
\begin{align}
	&\sigma_{Si} = 50\,, &&\sigma_{He} = 10^{-3}\,, && \sigma_{Au} = 10^4\,, && \sigma_{Ta} = 10^3 \,, \\
	&\rho c_{v,Si} = 20 \,, && \rho c_{v,He} = 0.01 \,, && \rho c_{v,Au} = 10^5 \,, && \rho c_{v,Ta} = 10^3  \,, \\
	&T_{hot,0} = 1 \,, && T_{cold,0}= 10^{-3} \,, && I_{d,hot,0} = B(T_{hot}) \,, && I_{d,cold,0} = B(T_{cold}) \,.%
\end{align} \label{eqns:material}%
\end{subequations}%
Units are $\mu$s for time, cm for space, 1 EU = $10^{12}$ erg for energy, and keV for temperature, yielding $a$ = 137.199 EU/cm$^4$/keV$^4$ and $c = 2.9979 \cdot 10^4$ cm/$\mu$s.
$\sigma$ and $\rho c_v$ in \refeqns{material} have units of cm$^{-1}$ and EU/cm$^3$/keV, respectively.
The starting time step is $6.25 \cdot 10^{-6}$ $\mu$s.
It grows by a constant geometric factor such that the problem time is $10^{-3}$ $\mu$s (or 1 ns) after \nth{30} time steps.
The time step is then fixed at $10^{-4}$ $\mu$s (or 0.1 ns) for the remainder of the problem.
For the finite element discretization, $I$, $\varphi$, $T$ consist of 2nd-order polynomials in $L_2$ while each component of $\boldsymbol{J}$ consists of 3rd-order polynomials in $H_1$.
An $S_{18}$ level-symmetric quadrature set is used for the angular discretization.
Four different simulations are performed: a reference simulation (10x more time steps), a ``fully implicit'' option in which the outer iteration of Algorithm~\ref{alg:TRTVEF} is fully converged, a ``one-sweep'' simulation (one transport sweep per time step), and a diffusion simulation performed by fixing $E = \frac{1}{3} I$.
The inner tolerance is $10^{-9}$ in all runs, and the outer tolerance is $10^{-8}$ in the ``fully implicit'' and reference runs.
All problems were run on 4 CPU nodes with 144 MPI ranks.
%

Figure~\ref{fig:results} shows $T$ at $t = 3$ ns (50 time steps) for the four different simulations.
Table~\ref{table:results} has total iteration counts at $t = 3$ ns, while Figure~\ref{fig:error} has relative errors in $T$.
The three curves with ``vs. $T_{ref}$'' in their labels in Figure~\ref{fig:error} show the $L_2$ error for $T$ relative to the reference result.
The fourth curve (``$T_{one}$ vs. $T_{full}$'') shows the difference between sweeping once per time step ($T_{one}$) and fully converging the outer iteration in Algorithm~\ref{alg:TRTVEF} ($T_{full}$).  
(Errors in $\varphi$ show similar trends and are omitted for brevity.)
More thorough analyses (more comparisons and problems) will be performed for the full paper.
Nonetheless, in the results here, we already see that the one-sweep option can produce a somewhat similar result to the fully implicit option, at a substantially lower cost.
Figures \ref{fig:results} and \ref{fig:error} show that difference between the one-sweep and fully implicit options is on the order of the temporal discretization error.
The larger difference in the 0.5-1.5 ns range indicates that it may be more desirable to do a few extra sweeps at this point in the simulation.
The difference between one-sweep and fully implicit approaches gets smaller in the later time steps.
Rather than choosing between performing 1 sweep and converging to the fully implicit solution, one could choose a looser outer iteration tolerance to obtain an intermediate method that falls between the two options in both cost and accuracy.

There are several additional findings beyond what we have shown in the figures and tables.
First, the solution diverges if we do not apply a negative flux correction on the transport system (this is true even if we redefine $E = \frac{1}{3} I$ in places with negative fluxes).
Second, we note that fully-upwind sweeps are used for these simulations; this is needed for the one-sweep approach.
If incoming fluxes are lagged across processor domain boundaries, the one-sweep approach has difficulties with convergence in this problem.
In such cases, it may be desirable for efficiency reasons to perform several outer iterations with lagged flux information, rather than to perform a single outer iteration with a fully-upwind sweep.
This tradeoff will be studied in future work.
Lastly, we note that the time-step ramping introduced for this problem is \emph{not} necessary for stability of the one-sweep (or fully implicit) schemes; it only serves to improve the solution accuracy.
Our experiments have indicated that the one-sweep method is stable for all the time steps we have considered.

\section{CONCLUSIONS}

We have presented a new approach for solving high-order TRT by leveraging the VEF equations.  
The scheme is robust even if only one transport sweep is performed per time step, but its outer iteration can be converged to attain the fully implicit TRT solution.
The one-sweep variant appears to be robust, preserves the TRT order of accuracy in time and space, minimizes the number of transport sweeps required to solve a TRT problem, and produces a reasonable solution that may be acceptable in many applications of interest.
In future work, we hope to study this algorithm on more problems and geometries, and we hope to extend the scheme to multigroup TRT problems.

%
%
\FloatBarrier 
\section*{ACKNOWLEDGEMENTS}

Livermore National Laboratory is operated by Lawrence Livermore National Security, LLC, for the U.S. Department of Energy, National Nuclear Security Administration under Contract DE-AC52-07NA27344.  This document (LLNL-PROC-820247) was prepared as an account of work sponsored by an agency of the U.S. government.  Neither the U.S. government nor Lawrence Livermore National Security, LLC, nor any of their employees makes any warranty, expressed or implied, or assumes any legal liability or responsibility for the accuracy, completeness, or usefulness of any information, apparatus, product, or process disclosed, or represents that its use would not infringe privately owned rights.  Reference herein to any specific commercial product, process, or service by trade name, trademark, manufacturer, or otherwise does not necessarily constitute or imply its endorsement, recommendation, or favoring by the U.S. government or Lawrence Livermore National Security, LLC. The views and opinions of authors expressed herein do not necessarily state or reflect those of the U.S. government or Lawrence Livermore National Security, LLC, and shall not be used for advertising or product endorsement purposes.

\setlength{\baselineskip}{12pt}
\bibliographystyle{mc2021}
\bibliography{mc2021}

\begin{thebibliography}{10}
\newcommand{\enquote}[1]{``#1''}
\providecommand{\url}[1]{\texttt{#1}}
\providecommand{\urlprefix}{URL }

\bibitem{langer2015performance}
S.~Langer, I.~Karlin, V.~Dobrev, M.~Stowell, and M.~Kumbera.
\newblock \enquote{Performance Analysis and Optimization for {BLAST}, a High
  Order Finite Element Hydro Code.}
\newblock Technical report, Lawrence Livermore National Lab (LLNL), Livermore,
  CA (United States) (2015).

\bibitem{dobrev2012high}
V.~A. Dobrev, T.~V. Kolev, and R.~N. Rieben.
\newblock \enquote{High-order curvilinear finite element methods for
  {L}agrangian hydrodynamics.}
\newblock \emph{SIAM Journal on Scientific Computing}, \textbf{volume~34}(5),
  pp. B606--B641 (2012).

\bibitem{anderson2018high}
R.~W. Anderson, V.~A. Dobrev, T.~V. Kolev, R.~N. Rieben, and V.~Z. Tomov.
\newblock \enquote{High-order multi-material {ALE} hydrodynamics.}
\newblock \emph{SIAM Journal on Scientific Computing}, \textbf{volume~40}(1),
  pp. B32--B58 (2018).

\bibitem{brunner2020nonlinear}
T.~A. Brunner, T.~S. Haut, and P.~F. Nowak.
\newblock \enquote{Nonlinear Elimination Applied to Radiation Diffusion.}
\newblock \emph{Nuclear Science and Engineering}, pp. 1--13 (2020).

\bibitem{olivier2017variable}
S.~S. Olivier and J.~E. Morel.
\newblock \enquote{Variable Eddington factor method for the {SN} equations with
  lumped discontinuous Galerkin spatial discretization coupled to a
  drift-diffusion acceleration equation with mixed finite-element
  discretization.}
\newblock \emph{Journal of Computational and Theoretical Transport},
  \textbf{volume~46}(6-7), pp. 480--496 (2017).

\bibitem{gol1964quasi}
V.~Y. Gol'Din.
\newblock \enquote{A quasi-diffusion method of solving the kinetic equation.}
\newblock \emph{USSR Computational Mathematics and Mathematical Physics},
  \textbf{volume~4}(6), pp. 136--149 (1964).

\bibitem{yee2020quadratic}
B.~C. Yee, S.~S. Olivier, T.~S. Haut, M.~Holec, V.~Z. Tomov, and P.~G. Maginot.
\newblock \enquote{A quadratic programming flux correction method for
  high-order DG discretizations of SN transport.}
\newblock \emph{Journal of Computational Physics} (2020).

\bibitem{brunner2006development}
T.~A. Brunner.
\newblock \enquote{Development of a grey nonlinear thermal radiation diffusion
  verification problem.}
\newblock Technical report, Sandia National Laboratory, Albuquerque, NM (2006).

\bibitem{olivier2021mc}
S.~S. Olivier, T.~S. Haut, and B.~C. Yee.
\newblock \enquote{Discontinuous Galerkin Variable Eddington Factor Methods.}
\newblock In \emph{Submitted to M\&C 2021}.

\bibitem{olivier2019high}
S.~S. Olivier, P.~G. Maginot, and T.~S. Haut.
\newblock \enquote{High Order Mixed Finite Element Discretization for the
  Variable Eddington Factor Equations.}
\newblock In \emph{Proceedings of the International Conference on Mathematics
  and Computational Methods applied to Nuclear Science and Engineering (M\&C
  2019)}. Portland, OR (2019).

\bibitem{holec2021mc}
M.~Holec, B.~S. Southworth, T.~S. Haut, W.~Pazner, and B.~C. Yee.
\newblock \enquote{Multi-group Nonlinear Diffusion Synthetic Acceleration of
  Thermal Radiative Transfer.}
\newblock In \emph{Submitted to M\&C 2021}.

\bibitem{fhr}
J.~Kallman, S.~MacLaren, K.~Baker, P.~Amala, K.~Lewis, and M.~Zika.
\newblock \enquote{{KULL} Simulations of {OMEGA} Radiation Flow Experiments.}
\newblock In \emph{American Physical Society Division of Plasma Physics
  Meeting} (2012).

\end{thebibliography}

\appendix
\gdef\thesection{APPENDIX \Alph{section}}
\section{Description of Material Boundaries for Half-Hohlraum Problem}
\label{app:a}

The purpose of this appendix is to precisely describe the locations of the four materials shown in Figure~\ref{fig:problem} for the half-hohlraum problem.
To simplify the description and avoid rounding errors, we will define locations in this abstract in units of $\Delta$ (see \refeqn{Delta}).
The Au wall is 6$\Delta$ thick (i.e., 6 spatial elements thick) around the outer domain boundary.
There are two regions of Si foam, each of which is 10$\Delta$ tall by 38$\Delta$ wide and flush with the right Au wall.
These two regions are centered symmetrically -- 23$\Delta$ from the top and bottom boundaries, respectively.
On each side of the Si foam, there is a Ta wall with thickness 2$\Delta$.
Each of these hohlraum regions is partially blocked in an asymmetric manner by an Au throttle spanning 2$\Delta$ in width.
The bottom throttle covers the upper Ta wall and 80\% of the Si foam region.
The top throttle covers the lower Ta wall and 20\% of the Si foam region.
The remainder of the domain is He.

\end{document}